\documentclass[journal,twocolumn]{IEEEtran}
\usepackage{}
\usepackage{graphicx}
\usepackage{epstopdf}
\usepackage{amssymb}
\usepackage{amsfonts}
\usepackage{amsmath}
\usepackage{algorithm}
\usepackage{algorithmic}
\usepackage{subeqnarray}
\usepackage{cases}
\usepackage{bm}
\usepackage{color}
\usepackage{subfigure,amsmath,amssymb,cite}
\usepackage{float}
\ifCLASSINFOpdf
\else
\fi
\begin{document}

\title{Joint Beamforming and Phase Shift Design for Hybrid-IRS-aided Directional Modulation Network}
\author{Rongen Dong, Hangjia He, Feng Shu, Riqing Chen, and Jiangzhou Wang,\emph{ Fellow, IEEE}

\thanks{This work was supported in part by the National Natural Science Foundation of China (Nos.U22A2002, and 62071234), the Major Science and Technology plan of Hainan Province under Grant ZDKJ2021022, and the Scientific Research Fund Project of Hainan University under Grant KYQD(ZR)-21008.}
\thanks{Rongen Dong and Feng Shu are with the School of Information and Communication Engineering, Hainan University, Haikou, 570228, China (Email: shufeng0101@163.com).}
\thanks{Hangjia He is with the School of Electronic and Optical Engineering, Nanjing University of Science and Technology, Nanjing, 210094, China.}
\thanks{Riqing Chen is with the Digital Fujian Institute of Big Data for Agriculture, Fujian Agriculture and Forestry University, Fuzhou 350002, China (Email: riqing.chen@fafu.edu.cn).}
\thanks{Jiangzhou Wang is with the School of Engineering, University of Kent, Canterbury CT2 7NT, U.K. (Email: {j.z.wang}@kent.ac.uk).}

%
}

\maketitle

\begin{abstract}
To make a good balance between performance, cost, and power consumption, a hybrid intelligent reflecting surface (IRS)-aided directional modulation (DM) network is investigated in this paper, where the hybrid IRS consists of passive and active reflecting elements. To maximize the achievable rate, two optimization algorithms, called maximum signal-to-noise ratio (SNR)-fractional programming (FP) (Max-SNR-FP) and maximum SNR-equal amplitude reflecting (EAR) (Max-SNR-EAR), are proposed to jointly design the beamforming vector and phase shift matrix (PSM) of hybrid IRS by alternately optimizing one and giving another. The former employs the successive convex approximation and FP methods to derive the beamforming vector and hybrid IRS PSM, while the latter adopts the maximum signal-to-leakage-noise ratio method and the criteria of phase alignment and EAR to design them. Simulation results show that the rates harvested by the proposed two methods are slightly lower than those of active IRS with higher power consumption, which are 35 percent higher than those of no IRS and random phase IRS, while passive IRS achieves only about 17 percent rate gain over the latter.
Moreover, compared to Max-SNR-FP, the proposed Max-SNR-EAR method makes an obvious complexity degradation at the price of a slight performance loss.

\end{abstract}
\section{Introduction}

Directional modulation (DM) has evolved into a useful strategy for fifth-generation millimeter-wave communication system, which can significantly boost the rate of wireless communication system\cite{Cheng2021Physical}. The design of DM synthesis is mainly carried out at the radio frequency frontend or baseband. For example, in \cite{Daly2009Directional}, the signal was generated in a predetermined direction through tuning the phase of each antenna element at the radio frequency frontend. In \cite{Shu2016Robust}, based on the maximizing signal-to-artificial noise (AN) ratio and maximizing signal-to-leakage-noise ratio methods designed at baseband, the AN projection matrix and precoder vector were obtained to maximize the secure rate of a multi-beam DM network.

Intelligent reflecting surface (IRS), proven to be an energy and cost-efficient tool for enhancing the performance of the wireless communication system, has been employed to assist various wireless communication scenarios: unmanned aerial vehicle communication \cite{Pan2021UAV}, single-cell wireless communication \cite{Wu2019Intelligent}, multi-cell communication \cite{Pan2020Multicell}, etc. Recently, IRS-assisted DM systems have also been investigated. In \cite{ShuEnhanced2021}, to maximize the secure rate of IRS-assisted DM network, the precoder vectors and phase shift matrix (PSM) of IRS were jointly devised by a high-performance general alternating iterative and low-complexity null-space projection algorithms.  For maximizing the receive power sum of IRS-assisted DM system, the authors in \cite{Dong2022Low} proposed the general alternating optimization and zero-forcing methods to jointly design the PSM at IRS and receive beamforming vectors at user.

However, all the above work was done based on fully passive IRS, and a satisfactory achievable rate of the system may not be ensured due to the effect of ``double fading'' in the cascaded channels. To effectively combat this effect and enhance the performance of the passive IRS-aided wireless communication network, the fully active IRS has been investigated \cite{Zhang2021Active, Liu2022Active}. However, the higher rate achieved by active IRS comes at the price of high hardware cost and power consumption. To overcome the limitations of fully passive and active IRSs, a hybrid active-passive IRS was proposed \cite{Nguyen2022Hybrid2, Nguyen2022Hybrid}. The main idea of the hybrid IRS is employing some active elements to substitute the one of the passive IRS, these active elements with signal amplification of hybrid IRS can effectively make up for the cascade path loss (PL) and increase the achievable rate. As far as the authors know, the hybrid IRS-aided DM system have not been investigated yet. In this article, we employ the hybrid IRS to further enhance the performance of passive IRS-aided DM network. The main contributions of this work are summarized as follows:
\begin{enumerate}
\item To make a good balance between performance, cost, and power consumption, a hybrid IRS-aided DM system model is proposed. Aiming at maximizing the achievable rate, the optimization problem of maximizing the signal-to-noise ratio (SNR) is established,
    and the maximum SNR-fractional programming (FP) (Max-SNR-FP) method is proposed to jointly optimize the beamforming vector and hybrid IRS PSM by solving one and giving another. In this scheme, the beamforming vector and passive IRS PSM are obtained by the successive convex approximation algorithm, and the active IRS PSM is computed by the FP method.

\item Given the high computational complexity of the Max-SNR-FP scheme, a low-complexity method, named maximum SNR-equal amplitude reflecting (EAR) (Max-SNR-EAR), is proposed. By utilizing the maximum signal-to-leakage-noise ratio (SLNR) method, the beamforming vector is obtained. Moreover, the hybrid IRS PSM is computed based on the criteria of phase alignment and EAR. From the simulation results, it is clear that the achievable rates harvested by both the proposed methods are higher than those of no IRS, random phase IRS, and passive IRS. In addition, when the number of hybrid IRS phase shift elements tends to be large, the difference in achievable rates between these two methods is trivial.
\end{enumerate}

The remainder of our work is organized as follows. In Section \ref{s1}, we describe the system model of hybrid IRS-aided DM network.
Section \ref{s2} presents the Max-SNR-FP scheme.
The Max-SNR-EAR scheme is described in Section \ref{s3}. We show the numerical simulation results in Section \ref{s4}. In Section \ref{s5}, the conclusions are drawn.

{\bf Notations:} in this article, the vectors and matrices are shown in boldface lowercase and uppercase letters, respectively. Symbols $(\cdot)^*$, $(\cdot)^T$, $(\cdot)^H$, Tr$(\cdot)$, $\Re\{\cdot\}$, and $\text{diag}\{\cdot\}$ stand for the conjugate, transpose, conjugate transpose, trace, real part, and diagonal operations, respectively. The sign $|\cdot|$ refers to the scalar's absolute value or the matrix's determinant. The notations $\textbf{I}_N$ and $\mathbb{C}^{N\times N}$ stand for the identity matrix and complex-valued matrix space of $N\times N$, respectively.

\vspace{-0.18mm}
\section{system model}\label{s1}
\begin{figure}[htbp]
\centering
\includegraphics[width=0.38\textwidth]{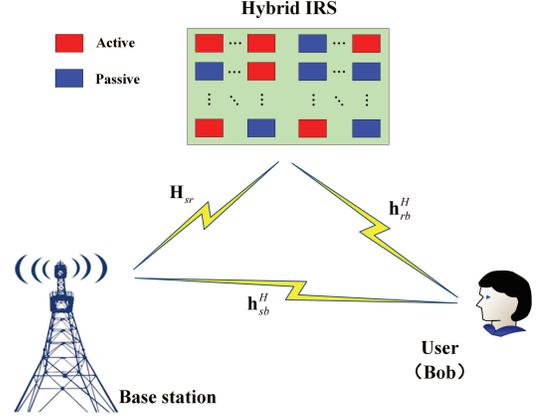}\\
\caption{System diagram of hybrid IRS-aided directional modulation network.}\label{model}
\end{figure}

As presented in Fig.~\ref{model}, a hybrid IRS-aided DM network is taken into account, where the base station (BS) and user (Bob) are equipped with $N$ antennas and single antenna, respectively. The hybrid IRS is equipped with $M$ elements, which consists of $M_a$ active and $M_p$ passive IRS reflecting elements ($1\leq M_a\leq M_p, M=M_a+M_p$). It is assumed that the active elements enable adjust both the amplitude and phase while the passive ones only tunes the phase of the incident signal. The signals reflected more than or equal to twice on the hybrid IRS are negligible due to the severe PL\cite{Pan2020Multicell}. All channels are assumed to be line-of-sight channels since DM is only applicable to such channels. We suppose that all the channel state information is completely accessible owing to the channel estimation\cite{Wang2020Channel}.

Similar to the conventional fully passive IRS, it is assumed that each elements of hybrid IRS can independently reflect the incident signals. Let us denote the set of the $M_a$ active elements by $\Omega$. $\boldsymbol{\Theta}=\text{diag}\{\boldsymbol{\theta}^*\}=\text{diag}\{\theta_1, \cdots, \theta_m, \cdots, \theta_M\}\in \mathbb{C}^{M\times M}$, $\boldsymbol{\Psi}=\text{diag}\{\boldsymbol{\psi}^*\}\in \mathbb{C}^{M\times M}$, and $\boldsymbol{\Phi}=\text{diag}\{\boldsymbol{\phi}^*\}\in \mathbb{C}^{M\times M}$ represent the reflection coefficients of total elements, active elements, and passive elements of hybrid IRS, respectively, where
\begin{eqnarray}
\theta_m=
\left\{
\begin{array}{ll}
\ |\beta_m|e^{j\mu_m},\quad &\text{if}~ m\in \Omega,\\
\ e^{j\mu_m},\quad &\text{otherwise},
\end{array}
\right.
\end{eqnarray}
$\mu_m\in[0,2\pi)$ is the phase, and $|\beta_m|$ represents the amplifying coefficient, which is subject to the power of the IRS active elements.
Let us define
\begin{align}
\boldsymbol{\Psi}=\textbf{E}_{M_a}\boldsymbol{\Theta},~\boldsymbol{\Phi}={\textbf{E}}_{M_p}\boldsymbol{\Theta},
\end{align}
where
\begin{align}
\textbf{E}_{M_a}+{\textbf{E}}_{M_p}=\textbf{I}_M,~ \textbf{E}_{M_a}{\textbf{E}}_{M_p}=\textbf{0}_M,
\end{align}
the non-zero elements of the diagonal matrix $\textbf{E}_{M_a}\in\mathbb{C}^{M\times M}$ are unity whose positions are determined by $\Omega$.

The transmitted signal at BS is expressed as
\begin{align}
\textbf{s}=\sqrt{P}\textbf{v}x,
\end{align}  
where $P$ represents the transmit power, $\textbf{v}\in\mathbb{C}^{N\times 1}$ and $x$ are the beamforming vector and the information symbol satisfying $\textbf{v}^H\textbf{v}=1$ and $\mathbb{E}[\|x\|^2]=1$, respectively.

In the presence of the PL, the received signal at  Bob is given by
\begin{align}\label{y_b}
y_b&=(\sqrt{\rho_{srb}}\textbf{h}^H_{rb}\boldsymbol{\Theta}\textbf{H}_{sr}+
\sqrt{\rho_{sb}}\textbf{h}^H_{sb})\textbf{s}+
\sqrt{\rho_{rb}}\textbf{h}^H_{rb}\boldsymbol{\Psi}\textbf{n}_r+n_b\nonumber\\
&=\sqrt{P}(\sqrt{\rho_{srb}}\textbf{h}^H_{rb}\boldsymbol{\Psi}\textbf{H}_{sr}+
\sqrt{\rho_{srb}}\textbf{h}^H_{rb}\boldsymbol{\Phi}\textbf{H}_{sr}+
\sqrt{\rho_{sb}}\textbf{h}^H_{sb})\textbf{v}x\nonumber\\
&~~~+\sqrt{\rho_{rb}}\textbf{h}^H_{rb}\boldsymbol{\Psi}\textbf{n}_r+n_b,
\end{align}
where $\rho_{srb}=\rho_{sr}\rho_{rb}$ represents the synthetic PL coefficient of BS-to-IRS channel and IRS-to-Bob channel, $\rho_{sb}$ and $\rho_{rb}$ stand for the PL coefficient of BS-to-Bob channel and IRS-to-Bob channel, respectively. $\textbf{n}_r\sim\mathcal {C}\mathcal {N}(\textbf{0}, \sigma^2_{r}\textbf{I}_{M_a})$ and $n_b\sim\mathcal {C}\mathcal {N}(0, \sigma^2_{b})$ denote the complex additive white Gaussian noise at the $M_a$ active elements of the hybrid IRS and at Bob, respectively.
$\textbf{h}_{sb}\in\mathbb{C}^{N\times 1}$, $\textbf{h}_{rb}\in\mathbb{C}^{M\times 1}$, and $\textbf{H}_{sr}=\textbf{h}_{sr}\textbf{h}^H_{sr}\in\mathbb{C}^{M\times N}$ represent the BS-to-Bob, IRS-to-Bob, and BS-to-IRS channels, respectively. Let us define the channel $\textbf{h}_{tr}=\textbf{h}(\theta_{tr})$,
the normalized steering vector $\textbf{h}(\theta)$ is expressed as
\begin{align}\label{h_theta}
\textbf{h}(\theta)\buildrel \Delta \over=\frac{1}{\sqrt{N}}[e^{j2\pi\Phi_{\theta}(1)}, \dots, e^{j2\pi\Phi_{\theta}(n)}, \dots, e^{j2\pi\Phi_{\theta}(N)}]^T,
\end{align}
where
\begin{align}
\Phi_{\theta}(n)\buildrel \Delta \over =-\left(n-\frac{N+1}{2}\right)\frac{d \cos\theta}{\lambda}, n=1, \dots, N,
\end{align}
$n$ stands for the antenna index, $d$ represents the spacing of adjacent transmitting antennas, $\theta$ means the direction angle of departure or arrival, and $\lambda$ denotes the wavelength.

In accordance with (\ref{y_b}), the achievable rate at Bob can be formulated as
\begin{align}\label{R_b}
&R_b=\log_2\left(1+\text{SNR}\right),
\end{align}
where
\begin{align}
\text{SNR}=\frac{P|(\sqrt{\rho_{srb}}\textbf{h}^H_{rb}\boldsymbol{\Psi}\textbf{H}_{sr}+
\sqrt{\rho_{srb}}\textbf{h}^H_{rb}\boldsymbol{\Phi}\textbf{H}_{sr}+
\sqrt{\rho_{sb}}\textbf{h}^H_{sb})\textbf{v}|^2}
{\sigma_r^2|\sqrt{\rho_{rb}}\textbf{h}^H_{rb}\boldsymbol{\Psi}|^2+\sigma_b^2}.
\end{align}
The transmit power of all active elements at the hybrid IRS is given by
\begin{align}\label{pr0}
P_r=\text{Tr}\left(\boldsymbol{\Psi}\Big(\rho_{sr} P\textbf{H}_{sr}\textbf{v}\textbf{v}^H\textbf{H}^H_{sr}+
\sigma_r^2\textbf{I}_{M}\Big)\boldsymbol{\Psi}^H\right),
\end{align}
which satisfies $P_r\leq P^{\text{max}}_r$, where $P^{\text{max}}_r$ represents the maximum transmit power of $M_a$ active elements.

In this work, we maximize the SNR by jointly optimizing beamforming vector $\textbf{v}$, passive IRS PSM $\boldsymbol{\Phi}$, and active IRS PSM $\boldsymbol{\Psi}$. The overall optimization problem is formulated as follows
\vspace{-1.5mm}
\begin{subequations}
\begin{align}\label{p0}
&\max \limits_{\textbf{v}, \boldsymbol{\Phi}, \boldsymbol{\Psi}}
~~\text{SNR}\\
&~~\text{s.t.} ~~~\textbf{v}^H\textbf{v}=1, P_r\leq P^{\text{max}}_r, \label{P_r}\\
& ~~~~~~~~|\boldsymbol{\Phi}(m,m)|=1, \text{if}~ m\not\in \Omega,\\
&~~~~~~~~|\boldsymbol{\Phi}(m,m)|=0, \text{otherwise},\label{phi5}\\
& ~~~~~~~~|\boldsymbol{\Psi}(m,m)|\leq \beta_{\text{max}}, \text{if}~m\in \Omega, \label{psi51}\\
&~~~~~~~~|\boldsymbol{\Psi}(m,m)|=0, \text{otherwise}, \label{psi5}
\end{align}
\end{subequations}
where $\beta_{\text{max}}$ is the amplitude budget. Considering that this optimization problem is a non-convex problem with a constant modulus constraint, and it is challenging to tackle it directly in general. In what follows, the alternating optimization algorithm is proposed to compute the beamforming vector and hybrid IRS PSM, respectively.

\section{Proposed Max-SNR-FP scheme}\label{s2}
In this section, we construct a Max-SNR-FP algorithm to jointly optimize the beamforming vector $\textbf{v}$, passive IRS PSM $\boldsymbol{\Phi}$, and active IRS PSM $\boldsymbol{\Psi}$. In what follows, we will alternately solve for $\textbf{v}$, $\boldsymbol{\Phi}$, and $\boldsymbol{\Psi}$.

\subsection{Optimize $\textbf{v}$ given $\boldsymbol{\Phi}$ and $\boldsymbol{\Psi}$}
Firstly, we transform the power constraint in (\ref{P_r}) into a convex constraint with respect to $\textbf{v}$ as follows
\begin{align}\label{p_r_v}
P_r=\textbf{v}^H\left(\rho_{sr} P \textbf{H}^H_{sr}\boldsymbol{\Psi}^H\boldsymbol{\Psi}\textbf{H}_{sr}\right)\textbf{v}+
\text{Tr}\left(\sigma_r^2\boldsymbol{\Psi}\boldsymbol{\Psi}^H\right)\leq P^{\text{max}}_r.
\end{align}
Then, given $\boldsymbol{\Phi}$ and $\boldsymbol{\Psi}$, the optimal beamforming vector $\textbf{v}$ can be found by addressing the problem in what follows
\begin{align}\label{w_u}
&~\max \limits_{\textbf{v}}
~\textbf{v}^H\textbf{A}\bar{\textbf{v}}~~~~~\text{s.t.}~\textbf{v}^H\textbf{v}=1,(\ref{p_r_v}),
\end{align}
where
\begin{align}
\textbf{A}=&(\sqrt{\rho_{srb}}\textbf{h}^H_{rb}\boldsymbol{\Phi}\textbf{H}_{sr}+
\sqrt{\rho_{srb}}\textbf{h}^H_{rb}\boldsymbol{\Psi}\textbf{H}_{sr}+\sqrt{\rho_{sb}}\textbf{h}^H_{sb})^H\nonumber\\
&(\sqrt{\rho_{srb}}\textbf{h}^H_{rb}\boldsymbol{\Phi}\textbf{H}_{sr}+
\sqrt{\rho_{srb}}\textbf{h}^H_{rb}\boldsymbol{\Psi}\textbf{H}_{sr}+\sqrt{\rho_{sb}}\textbf{h}^H_{sb}).
\end{align}
It is clear that this problem is not convex, and in accordance with the Taylor series expansion, we have
\begin{align}
\textbf{v}^H\textbf{A}\textbf{v}\geq 2\Re\{\bar{\textbf{v}}^H\textbf{A}\textbf{v}\}-\bar{\textbf{v}}^H\textbf{A}\bar{\textbf{v}},
\end{align}
where $\bar{\textbf{v}}$ is a given vector. Then (\ref{w_u}) can be recasted as
\begin{align}\label{w}
&~\max \limits_{\textbf{v}}
~2\Re\{\bar{\textbf{v}}^H\textbf{A}\textbf{v}\}-\bar{\textbf{v}}^H\textbf{A}\bar{\textbf{v}} ~~~~~\text{s.t.}~\textbf{v}^H\textbf{v}=1,(\ref{p_r_v}).
\end{align}
Given that this optimization problem is convex, we can obtain the optimal $\textbf{v}$ by adopting the CVX tool.
\subsection{Optimize $\boldsymbol{\Phi}$ given $\textbf{v}$ and  $\boldsymbol{\Psi}$}
In order to simplify the SNR expression with respect to the PSM $\boldsymbol{\Phi}$, we treat $\textbf{v}$ and  $\boldsymbol{\Psi}$ as two constants, and define
\begin{align}
B=(\sqrt{\rho_{srb}}\textbf{h}^H_{rb}\boldsymbol{\Psi}\textbf{H}_{sr}+
\sqrt{\rho_{sb}}\textbf{h}^H_{sb})\textbf{v}.
\end{align}
Then, the subproblem to optimize PSM $\boldsymbol{\Phi}$ is
\begin{subequations}\label{Phi_1}
\begin{align}
&~\max \limits_{\boldsymbol{\Phi}}
~|\sqrt{\rho_{srb}}\textbf{h}^H_{rb}\boldsymbol{\Phi}\textbf{H}_{sr}\textbf{v}+B|^2\\
&~~~\text{s.t.}~~|\boldsymbol{\Phi}(m,m)|=1,\text{if}~ m\not\in \Omega, \label{relax}\\
&~~~~~~~~|\boldsymbol{\Phi}(m,m)|=0, \text{otherwise}. \label{model0}
\end{align}
\end{subequations}
By defining
\begin{align}
\textbf{C}=\rho_{srb}\text{diag}\{\textbf{h}^H_{rb}\}\textbf{H}_{sr}\textbf{v}
\textbf{v}^H\textbf{H}^H_{sr}\text{diag}\{\textbf{h}^H_{rb}\}^H,
\end{align}
and based on the fact that diag$\{\textbf{p}\}\textbf{q}=\text{diag}\{\textbf{q}\}\textbf{p}$ for $\textbf{p}, \textbf{q}\in \mathbb{C}^{M\times 1}$, the objective function in (\ref{Phi_1}) can be recasted as
\begin{align}
\boldsymbol{\phi}^H\textbf{C}\boldsymbol{\phi}+
2\Re\{\sqrt{\rho_{srb}}\boldsymbol{\phi}^H\text{diag}\{\textbf{h}^H_{rb}\}\textbf{H}_{sr}\textbf{v}B^*\}+|B|^2.
\end{align}
According to the Taylor series expansion, we have
\begin{align}
\boldsymbol{\phi}^H\textbf{C}\boldsymbol{\phi}\geq 2\Re\{\bar{\boldsymbol{\phi}}^H\textbf{C}\boldsymbol{\phi}\}-\bar{\boldsymbol{\phi}}^H\textbf{C}\bar{\boldsymbol{\phi}},
\end{align}
where $\bar{\boldsymbol{\phi}}$ is a given vector. In addition, the unit modulus constraint (\ref{relax}) can be relaxed to
\begin{align}\label{relax1}
|\boldsymbol{\Phi}(m,m)|\leq1, \text{if}~ m\not\in \Omega.
\end{align}
At this point, the subproblem (\ref{Phi_1}) can be rewritten as follows
\begin{align}\label{Phi_2}
&~\max \limits_{\boldsymbol{\Phi}}
~2\Re\{\bar{\boldsymbol{\phi}}^H\textbf{C}\boldsymbol{\phi}\}-\bar{\boldsymbol{\phi}}^H\textbf{C}\bar{\boldsymbol{\phi}}
+|B|^2+2\Re\{\sqrt{\rho_{srb}}\boldsymbol{\phi}^H \bullet\nonumber\\
&~~~~~~~~~\text{diag}\{\textbf{h}^H_{rb}\}\textbf{H}_{sr}\textbf{v}B^*\}
~~~~~\text{s.t.}~~~ (\ref{relax1}), (\ref{model0}).
\end{align}
This problem can be solved directly with the CVX tool since it is convex.

\subsection{Optimize $\boldsymbol{\Psi}$ given $\textbf{v}$ and  $\boldsymbol{\Phi}$}
To optimize $\boldsymbol{\Psi}$, we regard $\textbf{v}$ and  $\boldsymbol{\Phi}$ as two given constants, and transform  the power constraint in (\ref{P_r}) into a convex constraint on $\boldsymbol{\psi}$ as follows
\begin{align}\label{p_max}
P_r&=\text{Tr}\left(\boldsymbol{\Psi}\Big(\rho_{sr} P\textbf{H}_{sr}\textbf{v}\textbf{v}^H\textbf{H}^H_{sr}+
\sigma^2\textbf{I}_{M}\Big)\boldsymbol{\Psi}^H\right)\nonumber\\
&=\boldsymbol{\psi}^T(\rho_{sr} P \text{diag}\{\textbf{v}^H\textbf{H}^H_{sr}\}
\text{diag}\{\textbf{H}_{sr}\textbf{v}\}+\sigma_r^2\textbf{I}_{M})\boldsymbol{\psi}^* \nonumber\\
&\leq P^{\text{max}}_r .
\end{align}
By neglecting the constant terms, the subproblem with respect to $\boldsymbol{\Psi}$ is given by
\begin{subequations}\label{Psi_1}
\begin{align}
&~\max \limits_{\boldsymbol{\Psi}}
~\frac{|(\sqrt{\rho_{srb}}\textbf{h}^H_{rb}\boldsymbol{\Psi}\textbf{H}_{sr}+
\sqrt{\rho_{srb}}\textbf{h}^H_{rb}\boldsymbol{\Phi}\textbf{H}_{sr}+
\sqrt{\rho_{sb}}\textbf{h}^H_{sb})\textbf{v}|^2}
{\sigma_r^2|\sqrt{\rho_{rb}}\textbf{h}^H_{rb}\boldsymbol{\Psi}|^2+\sigma_b^2}\\
&~~~\text{s.t.}~~~(\ref{psi51}), (\ref{psi5}), (\ref{p_max}).
\end{align}
\end{subequations}
Let us define
\begin{align}
D=(\sqrt{\rho_{srb}}\textbf{h}^H_{rb}\boldsymbol{\Phi}\textbf{H}_{sr}+
\sqrt{\rho_{sb}}\textbf{h}^H_{sb})\textbf{v}.
\end{align}
Then, the objective function in (\ref{Psi_1}) can be converted to
\begin{align}\label{Psi_3}
\frac{\boldsymbol{\psi}^H\textbf{C}\boldsymbol{\psi}+
2\Re\{\boldsymbol{\psi}^H\sqrt{\rho_{srb}}\text{diag}\{\textbf{h}^H_{rb}\}\textbf{H}_{sr}\textbf{v}D^*\}+|D|^2}
{\sigma_r^2\rho_{rb}|\boldsymbol{\psi}^H\text{diag}\{\textbf{h}^H_{rb}\}|^2+\sigma_b^2}.
\end{align}
At this point, the optimization problem (\ref{Psi_1}) has become a nonlinear fractional optimization problem. Based on  the FP strategy in \cite{Dinkelbach1967On}, we introduce a parameter $\tau$ and transform the objective function (\ref{Psi_3}) as
\begin{align}
&\boldsymbol{\psi}^H\textbf{C}\boldsymbol{\psi}+
2\Re\{\boldsymbol{\psi}^H\sqrt{\rho_{srb}}\text{diag}\{\textbf{h}^H_{rb}\}\textbf{H}_{sr}\textbf{v}D^*\}+|D|^2
\nonumber\\
&-\tau(\sigma_r^2\rho_{rb}|\boldsymbol{\psi}^H\text{diag}\{\textbf{h}^H_{rb}\}|^2+\sigma_b^2).
\end{align}
The optimal solution can be achieved if and only if $\boldsymbol{\psi}^H\textbf{C}\boldsymbol{\psi}+
2\Re\{\boldsymbol{\psi}^H\sqrt{\rho_{srb}}\text{diag}\{\textbf{h}^H_{rb}\}\textbf{H}_{sr}\textbf{v}D^*\}+|D|^2
-\tau(\sigma_r^2\rho_{rb}|\boldsymbol{\psi}^H\text{diag}\{\textbf{h}^H_{rb}\}|^2+\sigma_b^2)=0$.
We linearize the $\boldsymbol{\psi}^H\textbf{C}\boldsymbol{\psi}$ by employing Taylor series expansion at a given vector $\bar{\boldsymbol{\psi}}$, the subproblem with respect to $\boldsymbol{\Psi}$ can be recasted as
\begin{align}\label{Psi_2}
&~\max \limits_{\boldsymbol{\Psi}, \tau}
~~2\Re\{\bar{\boldsymbol{\psi}}^H\textbf{C}\boldsymbol{\psi}\}-\bar{\boldsymbol{\psi}}^H\textbf{C}\bar{\boldsymbol{\psi}}+
2\Re\{\boldsymbol{\psi}^H\sqrt{\rho_{srb}}\text{diag}\{\textbf{h}^H_{rb}\}\bullet
\nonumber\\
&~~~~~~~~~\textbf{H}_{sr}\textbf{v}D^*\}+|D|^2-
\tau(\sigma_r^2\rho_{rb}|\boldsymbol{\psi}^H\text{diag}\{\textbf{h}^H_{rb}\}|^2+\sigma_b^2)
\nonumber\\
&~~~\text{s.t.}~~~(\ref{psi51}), (\ref{psi5}), (\ref{p_max}).
\end{align}
Notice that problem (\ref{Psi_2}) is convex, and it can be addressed effectively with the CVX tool.
The whole procedure of the Max-SNR-FP scheme is described in Algorithm 1.
\begin{algorithm}
\caption{Proposed Max-SNR-FP algorithm}
\begin{algorithmic}[1]
\STATE Initialize feasible solutions $\textbf{v}^{(0)}$, $\boldsymbol{\Phi}^{(0)}$, and $\boldsymbol{\Psi}^{(0)}$, calculate achievable rate $R^{(0)}_b$ based on (\ref{R_b}).
\STATE Set the iteration number $k=0$, accuracy value $\epsilon$.
\REPEAT
\STATE Given $\boldsymbol{\Phi}^{(k)}$ and $\boldsymbol{\Psi}^{(k)}$, solve (\ref{w}) to get $\textbf{v}^{(k+1)}$.
\STATE Given $\textbf{v}^{(k+1)}$ and $\boldsymbol{\Psi}^{(k)}$, solve (\ref{Phi_2}) to get $\boldsymbol{\Phi}^{(k+1)}$.
\STATE Given $\textbf{v}^{(k+1)}$ and $\boldsymbol{\Phi}^{(k+1)}$, solve (\ref{Psi_2}) to get $\boldsymbol{\Psi}^{(k+1)}$.
\STATE Calculate $R^{(k+1)}_b$ based on $\textbf{v}^{(k+1)}$, $\boldsymbol{\Phi}^{(k+1)}$, and $\boldsymbol{\Psi}^{(k+1)}$.
\STATE Update $k=k+1$.
\UNTIL {$|R_b^{(k)}-R_b^{(k-1)}|\leq\epsilon$.}
\end{algorithmic}
\end{algorithm}

The overall computational complexity of the proposed Max-SNR-FP algorithm is
$\mathcal {O}(L((M+1)^3+2MN^2+2M^2)\text{In}(1/\epsilon)+M^3+N^3+5M^2+2MN+2M+2MN^2)$
float-point operations (FLOPs), where $L$ represents the numbers of alternating iterations, $\epsilon$ denotes the accuracy.


\vspace{-1mm}
\section{Proposed Max-SNR-EAR scheme}\label{s3}
In the previous section, we proposed the Max-SNR-FP method to compute the beamforming vector $\textbf{v}$, IRS phase shift matrices $\boldsymbol{\Phi}$ and $\boldsymbol{\Psi}$. However, it comes with a high computational complexity. To decrease the complexity, the Max-SNR-EAR scheme with low-complexity is proposed in this section.
\vspace{-2mm}
\subsection{Optimize $\textbf{v}$ given $\boldsymbol{\Phi}$ and $\boldsymbol{\Psi}$}
Given IRS phase shift matrices $\boldsymbol{\Phi}$ and $\boldsymbol{\Psi}$, in accordance with the principle of maximizing SLNR in \cite{Sadek2007A}, the beamforming vector $\textbf{v}$ can be optimized by tackling the problem as follows
\begin{align}\label{v2}
&~\max \limits_{\textbf{v}}
~\text{SLNR}=\frac{\textbf{v}^H\textbf{E}\textbf{v}}{\textbf{v}^H(\sigma_b^2\textbf{I}_N)\textbf{v}}
~~~~~\text{s.t.}~\textbf{v}^H\textbf{v}=1,(\ref{p_r_v}),
\end{align}
where
\begin{align}
\textbf{E}=&\rho_{srb}\textbf{H}_{sr}^H\boldsymbol{\Phi}^H\textbf{h}_{rb}\textbf{h}^H_{rb}\boldsymbol{\Phi}\textbf{H}_{sr}+
\rho_{srb}\textbf{H}_{sr}^H\boldsymbol{\Psi}^H\textbf{h}_{rb}\textbf{h}^H_{rb}\boldsymbol{\Psi}\textbf{H}_{sr}\nonumber\\
&+\textbf{h}_{sb}\textbf{h}^H_{sb}.
\end{align}
According to the Taylor series expansion and neglecting the constant terms, the problem (\ref{v2}) can be recasted as
\begin{align}\label{w2}
&~\max \limits_{\textbf{v}}
~2\Re\{\bar{\textbf{v}}^H\textbf{E}\textbf{v}\}-\bar{\textbf{v}}^H\textbf{E}\bar{\textbf{v}}
~~~~~\text{s.t.}~~~\textbf{v}^H\textbf{v}=1,(\ref{p_r_v}),
\end{align}
which can be addressed directly via adopting the CVX tool.
\subsection{Optimize $\boldsymbol{\Phi}$ and  $\boldsymbol{\Psi}$ given $\textbf{v}$}
Given beamforming vector $\textbf{v}$, we consider to design the phase of hybrid IRS firstly. The confidential message received by Bob through the cascade path is expressed as
\begin{align}
P\rho_{srb}\textbf{h}_{rb}^H\boldsymbol{\Theta}\textbf{H}_{sr}\textbf{v}
\textbf{v}^H\textbf{H}^H_{sr}\boldsymbol{\Theta}^H\textbf{h}_{rb}.
\end{align}
To maximize the confidential message of the cascade path, the phase alignment strategy is employed to design the hybrid IRS phase $\widetilde{\boldsymbol{\theta}}$,  $\widetilde{\boldsymbol{\theta}}$ is given by
\begin{align}\label{w_theta}
\widetilde{\boldsymbol{\theta}}=[e^{(-i\text{arg}(\textbf{s}_1))}, \cdots, e^{(-i\text{arg}(\textbf{s}_M))}]^T,
\end{align}
where $\textbf{s}=\text{diag}\{\textbf{h}_{rb}^H\}\textbf{H}_{sr}\textbf{v}$, and $\textbf{s}_k$ is the $k$-th element of $\textbf{s}$.

Next, inspired by the amplitude design of fully active IRS in \cite{Zhang2021Active}, we suppose that all active elements of the hybrid IRS have the same amplitude. Based on the IRS power constraint in (\ref{P_r}), we have
\vspace{-1mm}
\begin{align}\label{Q}
|\beta|=\sqrt{{P^{\max}_r}/{Q}},
\end{align}
\vspace{-3mm}
where
\begin{align}
Q=&\text{Tr}(\widetilde{\boldsymbol{\theta}}^H(\rho_{sr}P\text{diag}\{\textbf{v}^H\textbf{H}^H_{sr}\textbf{E}_{M_a}\}
\text{diag}\{\textbf{v}^H\textbf{H}^H_{sr}\textbf{E}_{M_a}\}^H \nonumber\\
&+\sigma^2\textbf{E}_{M_a}\textbf{E}_{M_a})\widetilde{\boldsymbol{\theta}}).
\end{align}
Based on (\ref{w_theta}) and (\ref{Q}), we obtain the passive IRS PSM and active IRS PSM as follows
\begin{align}
\boldsymbol{\Phi}=\textbf{E}_{M_p}\text{diag}\{\widetilde{\boldsymbol{\theta}}\},~
\boldsymbol{\Psi}=|\beta|\textbf{E}_{M_a}\text{diag}\{\widetilde{\boldsymbol{\theta}}\}.
\end{align}

Similar to Algorithm 1, we calculate $\textbf{v}$, $\boldsymbol{\Phi}$, and  $\boldsymbol{\Psi}$ alternately until convergence, i.e., $|R_b^{(k)}-R_b^{(k-1)}|\leq\epsilon$.
The overall computational complexity of Max-SNR-EAR scheme is
$\mathcal {O}(K(4M^2+N^3+8N^2M+2MN)$
FLOPs, where $K$ is the numbers of alternating iterations.

\section{Simulation Results}\label{s4}
Simulation results are shown to examine the performance of two proposed schemes in this section. Simulation parameters are given as follows: $N=8$, $M=128$, $M_a=32$, $d=\lambda/2$, $\theta_{sr}=\pi/4$, $\theta_{sb}=\pi/3$, $d_{sr}=200$m, $d_{sb}=220$m, $\sigma^2_b=-70$dBm, $\sigma^2_r=2\sigma^2_b$, $P=25$dBm, $P_r^{\text{max}}=30$dBm. The PL at the distance $d_{ab}$ is given by $\rho(d_{ab})=\text{PL}_0-10\gamma\text{log}_{10}\frac{d_{ab}}{d_0}$, where $\gamma$ represents the PL exponent, and $\text{PL}_0=-30$dB represents the PL reference distance $d_0=1$m. The PL exponents of all channels are chosen as 2. The positions of the hybrid IRS active elements are chosen as $\Omega=\{1, \cdots, M_a\}$.

\begin{figure}[htbp]
\vspace{-4mm}
\setlength{\abovecaptionskip}{-5pt}
\setlength{\belowcaptionskip}{-10pt}
\centering
\includegraphics[width=0.4\textwidth]{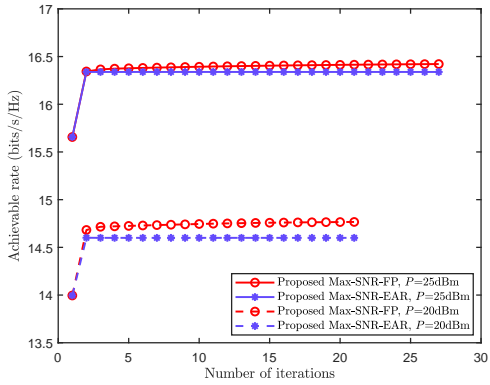}\\
\caption{Convergence of the proposed methods at different base station power.}\label{itea}
\vspace{-4.9mm}
\end{figure}

Firstly, the convergence behaviour of the proposed Max-SNR-FP and Max-SNR-EAR algorithms is investigated. Fig.~\ref{itea} presents the achievable rate versus the different base station power, i.e., $P=20$dBm, 25dBm. From Fig.~\ref{itea}, it can be seen that both of the proposed methods converge taking only a finite number of iterations. The proposed Max-SNR-EAR method has a faster convergence rate than the Max-SNR-FP method, regardless of $P=20$dBm or 25dBm.
	
\begin{figure}[htbp]
\vspace{-4mm}
\setlength{\abovecaptionskip}{-5pt}
\setlength{\belowcaptionskip}{-10pt}
\centering
\includegraphics[width=0.4\textwidth]{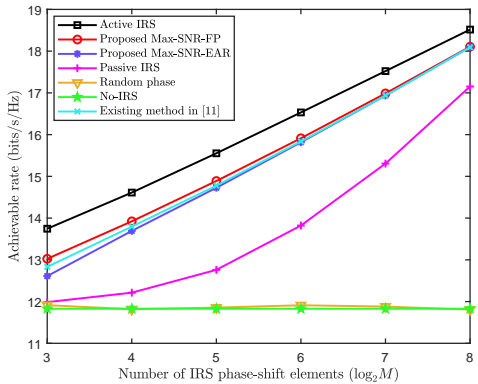}\\
\caption{Achievable rate versus the numbers of IRS phase shift elements.}\label{AR_m}
\vspace{-2mm}
\end{figure}

Fig.~\ref{AR_m} depicts the curves of the achievable rate versus the number of hybrid IRS phase shift elements $M$, where $M=2M_a$. We compare both proposed methods with the benchmark schemes: active IRS, passive IRS, no IRS, random phase IRS, and existing method in \cite{Nguyen2022Hybrid2}. The achievable rates of the proposed Max-SNR-FP and Max-SNR-EAR schemes gradually increase with the increases of $M$, and the former is better than the latter and existing method in \cite{Nguyen2022Hybrid2}. The achievable rates of both the proposed schemes outperform those of the passive IRS, random phase IRS, and no IRS. In addition, when $M$ tends to large-scale, the difference in achievable rates between both the proposed schemes and active IRS gradually decreases. 

\begin{figure}[htbp]
\vspace{-4mm}
\setlength{\abovecaptionskip}{-5pt}
\setlength{\belowcaptionskip}{-10pt}
\centering
\includegraphics[width=0.4\textwidth]{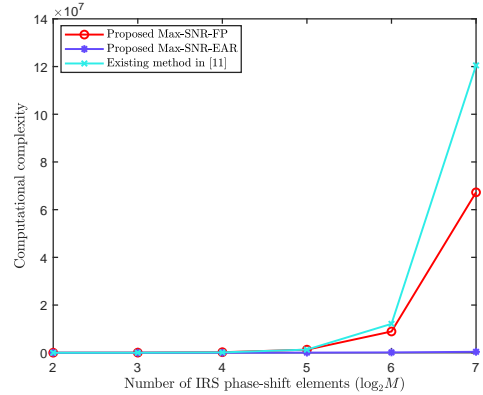}\\
\caption{Computational complexity versus the numbers of IRS elements.}\label{complex}
\vspace{-4.9mm}
\end{figure}

Fig.~\ref{complex} plots the curves of the computational complexity versus the number of hybrid IRS elements $M$. It can be found that the computational complexities of the proposed Max-SNR-FP method, proposed Max-SNR-EAR method, and existing method in \cite{Nguyen2022Hybrid2} are similar at small-scale IRS. However, when $M$ tends to large-scale, the complexities of the existing method in \cite{Nguyen2022Hybrid2} and proposed Max-SNR-FP method are far higher than those of the proposed Max-SNR-EAR method.

\vspace{-1mm}
\section{Conclusion}\label{s5}
In this work, we have made an investigation of the hybrid IRS-assisted DM network. To fully explore the advantages of hybrid IRS and maximize the achievable rate, the Max-SNR-FP and Max-SNR-EAR methods were proposed to jointly design the beamforming vector, passive IRS PSM, and active IRS PSM by alternately optimizing one and giving rest. From the simulation results, it can be found that the achievable rate of both proposed methods increases with the number of hybrid IRS elements increases, and is superior to those of no IRS, random phase IRS, and passive IRS. Moreover, the proposed Max-SNR-FP method exceeds the existing method with respect to the achievable rate and has lower complexity.



\ifCLASSOPTIONcaptionsoff
  \newpage
\fi

\bibliographystyle{IEEEtran}
\vspace{-1.8mm}
\bibliography{IEEEfull,reference}
\end{document}